\documentclass{article}
\usepackage{amsfonts}
\usepackage{amssymb}
\usepackage{amsmath}
\usepackage{amsthm}
\usepackage[english]{babel}
\usepackage{epigraph}

\begin{document}

\title{Static spherically symmetric solutions\\
in New General Relativity}

\author{Alexey Golovnev$^1$, A. N. Semenova$^2$, V.P. Vandeev$^2$\\
{\small \it $^1$Centre for Theoretical Physics, The British University in Egypt,}\\
{\small \it El Sherouk City, Cairo 11837, Egypt}\\
{\small  agolovnev@yandex.ru}\\
{\small \it $^2$Petersburg Nuclear Physics Institute of National Research Centre ``Kurchatov Institute'',}\\ 
{\small \it Gatchina, 188300, Russia}\\
{\small ala.semenova@gmail.com \hspace{50 pt} vandeev{\_}vp@pnpi.nrcki.ru} } 
\date{}

\maketitle

\begin{abstract}

We give a pedagogical introduction to static spherically symmetric solutions in models of New GR, both explaining the basics and showing how all such vacuum solutions can be obtained in elementary functions. In doing so, we coherently introduce the full landscape of  these modified teleparallel spacetimes, and find a few special cases. The equations of motion are turned into a very nice and compact form by using the Levi-Civita divergence of the torsion-conjugate; and generalised Bianchi identities are briefly discussed. Another important point we make  is that a convenient choice of the radial variable might be instrumental for success of similar studies in other modified gravity models.

\end{abstract}

\section{Introduction}

Given the well-known troubles of modern precision cosmology, various approaches to modified gravity are very popular with researchers. There are many ways of modifying it, with variable degree of problems and difficulties. One path which is clearly on the rise these days is the teleparallel framework in terms of torsion \cite{revtor}. There are many versions of teleparallel modifications on the market, with their own deep issues of viability \cite{issues}. Unfortunately, we often march towards phenomenological applications without dwelling much on the foundational properties.

And even apart from the foundational issues, many modified models of teleparallel type appear to be much more complicated than it has ever been expected. For instance, we haven't been able to find non-trivial exact Black Hole solutions in $f(\mathbb T)$ gravity, unless we accept topological flat horizons \cite{notspher} or a complex tetrad \cite{morespher} which is problematic since the tetrad is the dynamical variable in teleparallel gravity. Therefore it definitely makes sense to better study the physical  properties of the simplest models.

One of the simplest modified gravity ideas we can think of, in the teleparallel realm, is the New GR theory \cite{HaSh} proposed a long time ago. In this paper, we set out to give a pedagogical introduction to static spherically symmetric configurations in these models, with the main message being that in vacuum absolutely any such solution, in an arbitrary model of this type, can be found in elementary functions. We explicitly show how it can be done.

We have to note that a restriction on model parameters was imposed in the Ref. \cite{HaSh} from the viewpoint of  Newtonian limit, in the form of the first $\frac{1}{r}$-correction to the zero Newtonian potential at infinity. More precisely, given the mass of the central body, and assuming that gravity is everywhere weak, they wanted that its coefficient corresponded to the usual one for GR, that is Newtonian gravity. 

However, for the vacuum solutions, this coefficient is just an integration constant, while ascribing it to the mass which ended up in the central singularity is a matter of interpretation. And the real discussion of the Newtonian limit would require more than the static spherically symmetric solutions alone. On the other hand, changing this coefficient can be reduced to introducing an overall factor for the gravitational part of the action, i.e. rescaling the gravitational constant, and therefore mathematically it doesn't modify the behaviour of solutions at all. Moreover, changes of the gravitational constant might finally come up from effects of a non-linear modification of the New GR. For all these reasons, we will look for solutions without restricting the model parameters anyhow.

The paper is organised as follows. We first briefly summarise the basics of the Teleparallel Equivalent of General Relativity (TEGR), then introduce the generalised models of New GR, explain the derivation of their equations of motion in detail, and finally discuss all the static spherically symmetric solutions thereof.

\section{A brief reminder of TEGR}

Teleparallel theories are those metric-affine approaches to gravity which use purely flat (no curvature) connections. Roughly speaking, they can be of two different kinds: metric teleparallel in terms of torsion, or symmetric teleparallel in terms of non-metricity. In this paper, we follow the former. In particular, we assume that, given a metric $g_{\mu\nu}$, there exists an orthonormal
$$g_{\mu\nu}=\eta_{ab}e^a_{\mu} e^b_{\nu}$$
set of 1-forms $e^a_{\mu}$, with the dual basis of vectors $e^{\mu}_a$ given by the matrix inverse of $e^a_{\mu}$, such that they are covariantly constant
$$\bigtriangledown_{\mu} e^a_{\nu}=0, \qquad \bigtriangledown_{\mu} e_a^{\nu}=0$$
assuming no spin connection for that. This is the geometrical meaning of metric teleparallelism \cite{meGeom}.

Let us also mention right away that we follow what is called the pure-tetrad approach to metric teleparallel theories. Since we have chosen a basis of vectors to be covariantly constant over the whole manifold, this set cannot be arbitrarily changed without changing the very geometry behind, therefore the local Lorentz invariance is lost. Of course, as always, the broken symmetry can be formally restored by a kind of St{\"u}ckelberg trick. In teleparallel models, it is achieved by introducing a spin connection \cite{others}. However, the resulting model is no different from the pure-tetrad approach \cite{Lor}, and we prefer to stay with the latter. The Lorentz-covariant way of dealing with it can be found, for instance, in the Ref. \cite{Obukhov}.

All in all, we have a basis of covariantly constant 1-forms $e^a_{\mu}$, with the Latin index just numbering the 1-forms. It is easy to see that the corresponding connection coefficient is then
\begin{equation}
\label{Weitz}
\Gamma^{\alpha}_{\mu\nu}=e^{\alpha}_a \partial_{\mu} e^a_{\nu},
\end{equation}
where our convention is 
$$\bigtriangledown_{\mu} A^{\nu} = \partial_{\mu} A^{\nu} + \Gamma^{\nu}_{\mu\alpha} A^{\alpha}$$ which has a position of lower indices different from many other works. This connection (\ref{Weitz}) is obviously flat and metric-compatible. However, it does have non-zero torsion,
\begin{equation}
\label{tor}
T^{\alpha}_{\hphantom{\alpha}\mu\nu} = \Gamma^{\alpha}_{\mu\nu} - \Gamma^{\alpha}_{\nu\mu}.
\end{equation}

One can find the general description of teleparallel gravity \cite{meCov} in many works. Let us only give a very brief summary of it here. It's a very standard and easy exercise, to  prove that any metric-compatible connection $\Gamma^{\alpha}_{\mu\nu}$ can be given in terms of the Levi-Civita one ${\mathop \Gamma\limits^{(0)}}{}^{\alpha}_{\mu\nu} $ as
\begin{equation}
\label{cont}
\Gamma^{\alpha}_{\mu\nu}={\mathop \Gamma\limits^{(0)}}{}^{\alpha}_{\mu\nu} + K^{\alpha}_{\hphantom{\alpha}\mu\nu}
= {\mathop \Gamma\limits^{(0)}}{}^{\alpha}_{\mu\nu} + \frac12 \left(T^{\alpha}_{\hphantom{\alpha}\mu\nu} + T^{\hphantom{\mu}\alpha}_{\mu\hphantom{\alpha}\nu} - T^{\hphantom{\nu\mu}\alpha}_{\nu\mu}  \right)
\end{equation}
with the contortion tensor $ K_{\alpha\mu\nu}=- K_{\nu\mu\alpha}$. Calculating the vanishing curvature tensor $R^{\alpha}_{\hphantom{\alpha}\beta\mu\nu}(\Gamma)=0$ in terms of $R^{\alpha}_{\hphantom{\alpha}\beta\mu\nu}({\mathop\Gamma\limits^{0}})$, we get
\begin{equation}
\label{basrel}
0={\mathop R\limits^{(0)}} + 2 {\mathop \bigtriangledown\limits^{(0)}}_{\mu} T^{\mu} + \mathbb T
\end{equation}
with the torsion vector and torsion scalar defined as
$$T_{\mu} = T^{\alpha}_{\hphantom{\alpha}\mu\alpha}\qquad \mathrm{and}\qquad {\mathbb T}= \frac12 S^{\alpha\mu\nu} T_{\alpha\mu\nu} =  \frac14  T^{\alpha\mu\nu} T_{\alpha\mu\nu} + \frac12 T^{\alpha\mu\nu} T_{\mu\alpha\nu} -  T^{\mu}T_{\mu}, $$
and the so-called superpotential or torsion-conjugate
$$ S^{\alpha\mu\nu} =  K^{\mu\alpha\nu} + g^{\alpha\mu}T^{\nu} - g^{\alpha\nu}T^{\mu}=\frac12\left(T^{\alpha\mu\nu} +T^{\mu\alpha\nu} -T^{\nu\alpha\mu} \right)+ g^{\alpha\mu}T^{\nu} - g^{\alpha\nu}T^{\mu}$$
with the same antisymmetry $S^{\alpha\mu\nu} =- S^{\alpha\nu\mu} $ as of the torsion tensor itself.

Given the relation between the two scalars (\ref{basrel}), and neglecting the boundary term, we see that the two actions,
$$S_{\mathrm{GR}}=-\int d^4 x \sqrt{-g}\cdot {\mathop R\limits^{(0)}} \qquad \mathrm{and} \qquad  S_{\mathrm{TEGR}}=\int d^4 x \sqrt{-g}\cdot \mathbb T$$
are equivalent, and therefore the latter is called Teleparallel Equivalent of General Relativity (TEGR).

\section{New General Relativity}

Now we turn to a generalisation of TEGR which was dubbed New GR \cite{HaSh}. For that, we define a generalisation of the usual torsion scalar
\begin{equation}
\label{torscal}
{\mathfrak T} = \frac12 {\mathfrak S}^{\alpha\mu\nu} T_{\alpha\mu\nu}=  \frac{a}{4}\cdot T^{\alpha\mu\nu} T_{\alpha\mu\nu} + \frac{b}{2}\cdot T^{\alpha\mu\nu} T_{\mu\alpha\nu} - c\cdot T^{\mu}T_{\mu},
\end{equation}
with the same torsion vector $T_{\mu}$ as in the previous Section, and the corresponding generalised torsion-conjugate ${\mathfrak S}^{\alpha\mu\nu} = -{\mathfrak S}^{\alpha\nu\mu} $,
\begin{equation}
\label{superpot}
{\mathfrak S}^{\alpha\mu\nu} = \frac{a}{2} T^{\alpha\mu\nu} +\frac{b}{2}\left(T^{\mu\alpha\nu} -T^{\nu\alpha\mu} \right)+c\left( g^{\alpha\mu}T^{\nu} - g^{\alpha\nu}T^{\mu}\right),
\end{equation}
which depend on the three constant numbers, $a,b,c$.  It is the most general parity-preserving quadratic scalar in terms of the torsion tensor, and our action will be
$$S_{\mathrm{NewGR}}=\int d^4 x \sqrt{-g}\cdot \mathfrak T = \frac12 \int d^4 x \sqrt{-g}\cdot {\mathfrak S}^{\alpha\mu\nu} T_{\alpha\mu\nu}.$$
Obviously, when $a=b=c=1$ we simply have ${\mathfrak T}={\mathbb T}$ and ${\mathfrak S}^{\alpha\mu\nu} =  S^{\alpha\mu\nu} $, and are back to TEGR. Any other common non-zero value of $a=b=c$ differs from GR by only a new value of the effective gravitational constant.

Note that the initial paper on New GR \cite{HaSh} used another representation of the same quadratic form. In particular, it had got a square of the torsion axial vector (the totally antisymmetric part of the torsion tensor). In our notations (\ref{torscal}) it corresponds to the first two terms with $b=-a$. As we will see below, the axial part of torsion is identically zero for the configurations we study, and therefore the $a$ and $b$ coefficients  will appear only in the combination of $a+b$. Of course, one can equivalently use any other convenient parametrisation of the torsion scalar \cite{varopt}.

Finally, in order to get prepared for deriving the equations of motion, note that the tetrad is assumed to be the only fundamental variable of the theory, and it influences the action through both the metric and the torsion tensor components. The variation of the metric can be done very easily as
\begin{equation}
\label{varmet}
\delta g_{\mu\nu} = \eta_{ab}\left( e^{a}_{\mu} \delta e^b_{\nu} +(\delta e^{a}_{\mu}) e^b_{\nu}\right) = 2 \eta_{ab} e^{a}_{\mu} \delta e^b_{\nu}
\end{equation}
$$ \mathrm{with} \qquad  \delta g^{\mu\nu}= - g^{\mu\alpha} g^{\nu\beta} \delta g_{\alpha\beta} \qquad \mathrm{and} \qquad \delta \sqrt{-g} = \frac12 \sqrt{-g}\cdot g^{\mu\nu} \delta g_{\mu\nu}.$$
Furthermore, the linear variation of the connection coefficient (\ref{Weitz}) can be presented in a nice form:
$$\delta \Gamma^{\alpha}_{\mu\nu}= e^{\alpha}_a \partial_{\mu} \delta e^a_{\nu} - e^{\alpha}_b e^{\beta}_a \delta e^b_{\beta}\cdot \partial_{\mu} e^a_{\nu}= e^{\alpha}_a \bigtriangledown_{\mu} \delta e^a_{\nu} =\bigtriangledown_{\mu} \left( e^{\alpha}_a  \delta e^a_{\nu}\right).$$
Specifying it to the torsion tensor (\ref{tor}), we get
\begin{equation}
\label{vartor}
\delta T^{\alpha}_{\hphantom{\alpha}\mu\nu}= \bigtriangledown_{\mu} \left( e^{\alpha}_a  \delta e^a_{\nu}\right) - \bigtriangledown_{\nu} \left( e^{\alpha}_a  \delta e^a_{\mu}\right).
\end{equation}
Of course, in order to effectively use these relations, we need to know how integration by parts works with our connection.

\subsection{Integration by parts with the teleparallel connection}

Let's assume we have a vector ${\mathcal V}^{\mu}$ and a tensor ${\mathcal T}_{\mu\nu}$, and transform the following expression, with an arbitrary metric-compatible connection (\ref{cont}) with torsion:
\begin{multline*}
\sqrt{-g} {\mathcal T}^{\mu\nu}\bigtriangledown_{\mu}{\mathcal V}_{\nu}= \sqrt{-g} {\mathcal T}^{\mu\nu}\left( \partial_{\mu}{\mathcal V}_{\nu} - \Gamma^{\alpha}_{\mu\nu}{\mathcal V}_{\alpha}\right)= \partial_{\mu} \left(\sqrt{-g} {\mathcal T}^{\mu\nu}{\mathcal V}_{\nu}\right) - {\mathcal V}_{\nu} \partial_{\mu} \left(\sqrt{-g} {\mathcal T}^{\mu\nu}\right)  - \sqrt{-g} {\mathcal T}^{\mu\nu} \Gamma^{\alpha}_{\mu\nu}{\mathcal V}_{\alpha}\\
= \partial_{\mu} \left(\sqrt{-g} {\mathcal T}^{\mu\nu}{\mathcal V}_{\nu}\right) - \sqrt{-g} {\mathcal V}_{\nu} \left(\partial_{\mu}{\mathcal T}^{\mu\nu}+{\mathop\Gamma^{(0)}}{}^{\alpha}_{\mu\alpha} {\mathcal T}^{\mu\nu} + \Gamma^{\nu}_{\mu\alpha}{\mathcal T}^{\mu\alpha}  \right)\\
=\partial_{\mu} \left(\sqrt{-g} {\mathcal T}^{\mu\nu}{\mathcal V}_{\nu}\right) - \sqrt{-g} {\mathcal V}_{\nu} \left( \bigtriangledown_{\mu}{\mathcal T}^{\mu\nu} - K^{\mu}_{\hphantom{\mu}\mu\alpha} {\mathcal T}^{\alpha\nu}\  \right) = \partial_{\mu} \left(\sqrt{-g} {\mathcal T}^{\mu\nu}{\mathcal V}_{\nu}\right) - \sqrt{-g} {\mathcal V}_{\nu} \left( \bigtriangledown_{\mu}{\mathcal T}^{\mu\nu} + T_{\alpha} {\mathcal T}^{\alpha\nu}\  \right).
\end{multline*}
Neglecting the boundary term, it proves the following relation:
\begin{equation}
\label{byparts}
\int d^4 x  \sqrt{-g} {\mathcal T}^{\mu\nu}\bigtriangledown_{\mu}{\mathcal V}_{\nu} = - \int d^4 x \sqrt{-g} {\mathcal V}_{\nu} \left( \bigtriangledown_{\mu} + T_{\mu}\right){\mathcal T}^{\mu\nu}.
\end{equation}
Obviously, having chosen any other ranks of the tensors would not change anything in this result.

\section{Derivation of equations of motion}

Now we have to vary the action. Using the variation of the metric (\ref{varmet}), one of the terms in the action variation is given simply by
\begin{equation}
\label{trvar}
\delta\sqrt{-g}\cdot \mathfrak T=  \frac12 \sqrt{-g}\cdot {\mathfrak T}\cdot g^{\mu\nu} \delta g_{\mu\nu}= \sqrt{-g}\cdot {\mathfrak T}\cdot g^{\mu\nu} \eta_{ab} e^{a}_{\mu} \delta e^b_{\nu} =  \sqrt{-g}\cdot {\mathfrak T}\delta^{\nu}_{\mu}\cdot e^{\mu}_a \delta e^a_{\nu},
\end{equation}
and of course it is a very standard part of the game. 

In order to facilitate the rest of the computations, let's define
$${\mathcal E}^{\hphantom{\alpha} \mu\nu \hphantom{\beta} \rho\sigma}_{\alpha \hphantom{\mu\nu} \beta} = \frac{a}{4}\cdot g_{\alpha\beta} g^{\mu\rho} g^{\nu\sigma} + \frac{b}{2}\cdot \delta^{\rho}_{\alpha} \delta^{\mu}_{\beta} g^{\nu\sigma} - c \cdot \delta^{\nu}_{\alpha} \delta^{\sigma}_{\beta} g^{\mu\rho} $$
so that
$${\mathfrak T} = {\mathcal E}^{\hphantom{\alpha} \mu\nu \hphantom{\beta} \rho\sigma}_{\alpha \hphantom{\mu\nu} \beta} T^{\alpha}_{\hphantom{\alpha}\mu\nu} T^{\beta}_{\hphantom{\beta}\rho\sigma},$$
and the variation is
\begin{equation}
\label{varT}
\delta {\mathfrak T} =\delta {\mathcal E}^{\hphantom{\alpha} \mu\nu \hphantom{\beta} \rho\sigma}_{\alpha \hphantom{\mu\nu} \beta}\cdot T^{\alpha}_{\hphantom{\alpha}\mu\nu} T^{\beta}_{\hphantom{\beta}\rho\sigma} + 2 {\mathcal E}^{\hphantom{\alpha} \mu\nu \hphantom{\beta} \rho\sigma}_{\alpha \hphantom{\mu\nu} \beta} T^{\alpha}_{\hphantom{\alpha}\mu\nu} \delta T^{\beta}_{\hphantom{\beta}\rho\sigma} 
\end{equation}
since ${\mathcal E}^{\hphantom{\alpha} \mu\nu \hphantom{\beta} \rho\sigma}_{\alpha \hphantom{\mu\nu} \beta} = {\mathcal E}^{\hphantom{\beta} \rho\sigma \hphantom{\alpha} \mu\nu}_{\beta \hphantom{\rho\sigma} \alpha} $.

In the variation (\ref{varT}) of $\mathfrak T$, the second term can be transformed as
\begin{multline*}
2 {\mathcal E}^{\hphantom{\alpha} \mu\nu \hphantom{\beta} \rho\sigma}_{\alpha \hphantom{\mu\nu} \beta} T^{\alpha}_{\hphantom{\alpha}\mu\nu}\cdot \delta T^{\beta}_{\hphantom{\beta}\rho\sigma} = 2\left( {\mathcal E}^{\hphantom{\alpha} \mu\nu \hphantom{\beta} \rho\sigma}_{\alpha \hphantom{\mu\nu} \beta} - {\mathcal E}^{\hphantom{\alpha} \mu\nu \hphantom{\beta} \sigma\rho}_{\alpha \hphantom{\mu\nu} \beta} \right) T^{\alpha}_{\hphantom{\alpha}\mu\nu}\cdot \bigtriangledown_{\rho}\left(e^{\beta}_a \delta e^a_{\sigma}\right) \\
=2 {\mathfrak S}_{\beta}^{\hphantom{\beta}\rho\sigma}\cdot \bigtriangledown_{\rho}\left(e^{\beta}_a \delta e^a_{\sigma}\right) =-  2 {\mathfrak S}_{\mu}^{\hphantom{\mu}\nu\alpha}\cdot \bigtriangledown_{\alpha}\left(e^{\mu}_a \delta e^a_{\nu}\right).
\end{multline*}
Upon integration by parts (\ref{byparts}), it produces the following contribution to the equations:
\begin{equation}
\label{simpvar}
2\sqrt{-g}\cdot {\mathcal E}^{\hphantom{\alpha} \mu\nu \hphantom{\beta} \rho\sigma}_{\alpha \hphantom{\mu\nu} \beta} T^{\alpha}_{\hphantom{\alpha}\mu\nu}\cdot \delta T^{\beta}_{\hphantom{\beta}\rho\sigma} \longrightarrow 2\sqrt{-g} \cdot \left(\bigtriangledown_{\alpha} + T_{\alpha} \right) {\mathfrak S}_{\mu}^{\hphantom{\mu}\nu\alpha}\cdot e^{\mu}_a \delta e^a_{\nu},
\end{equation}
the  one which contains an antisymmetric part since generically ${\mathfrak S}_{\mu\nu\alpha}$ is neither symmetric nor antisymmetric in its first two indices.

Finally, the first term of $\delta\mathfrak T$ variation (\ref{varT}) is easily calculated to be
\begin{equation}
\label{finvar}
\delta {\mathcal E}^{\hphantom{\alpha} \mu\nu \hphantom{\beta} \rho\sigma}_{\alpha \hphantom{\mu\nu} \beta}\cdot T^{\alpha}_{\hphantom{\alpha}\mu\nu} T^{\beta}_{\hphantom{\beta}\rho\sigma} = \left(\frac{a}{2}T^{\nu\alpha\beta}T_{\mu\alpha\beta} - a T^{\alpha\beta\nu}T_{\alpha\beta\mu} - b T^{\alpha\beta\nu}T_{\beta\alpha\mu}+ 2c T^{\nu}T_{\mu} \right) \cdot e^{\mu}_a \delta e^a_{\nu}.
\end{equation}
In principle, everything is ready for presenting the equations of motion, but we would like to rewrite this expression in a way independent of the arbitrary constants, by absorbing them directly into the torsion conjugate (\ref{superpot}). For doing so, one can check that
\begin{equation}
\label{thequant}
{\mathfrak S}_{\mu\alpha\beta} T^{\nu\alpha\beta} -  2 {\mathfrak S}^{\alpha\beta\nu} T_{\alpha\beta\mu} = \frac{a}{2}T^{\nu\alpha\beta}T_{\mu\alpha\beta} - a T^{\alpha\beta\nu}T_{\alpha\beta\mu} - b T^{\alpha\beta\nu}T_{\beta\alpha\mu}+ 2c T^{\nu}T_{\mu}, 
\end{equation}
precisely the variation we need. Note that this is a $\mu\leftrightarrow\nu$ symmetric tensor, as it must be for any variation with respect to the metric tensor.

There is a simple reason for why the result of this form actually comes out. The superpotential has been constructed in such a way as $T^{\hphantom{\alpha} \mu\nu}_{\alpha} \delta {\mathfrak S}_{\hphantom{\alpha} \mu\nu}^{\alpha}={\mathfrak S}^{\hphantom{\alpha} \mu\nu}_{\alpha} \delta T_{\hphantom{\alpha} \mu\nu}^{\alpha}$, and what is then left  in $\delta\mathfrak T$, after having taken the term (\ref{simpvar}), is the variation of the metric tensor in ${\mathfrak T}=g_{\alpha\beta} g^{\mu\rho} g^{\nu\sigma} {\mathfrak S}_{\hphantom{\alpha} \mu\nu}^{\alpha} T_{\hphantom{\beta} \rho\sigma}^{\beta}$. The only non-trivial fact is that the quantity (\ref{thequant}) is automatically $\mu\leftrightarrow\nu$ symmetric, and therefore does not require any extra symmetrisation.

Finally, summing all the contributions (\ref{trvar}, \ref{simpvar}, \ref{finvar}) up, the equations in vacuum take the following form:
\begin{equation}
\label{fulleq}
\left(\bigtriangledown_{\alpha} + T_{\alpha} \right) {\mathfrak S}_{\mu}^{\hphantom{\mu}\nu\alpha}+\frac12 {\mathfrak S}_{\mu\alpha\beta} T^{\nu\alpha\beta} -   {\mathfrak S}^{\alpha\beta\nu} T_{\alpha\beta\mu} + \frac12 {\mathfrak T}\delta^{\nu}_{\mu}=0. 
\end{equation}
which corresponds to how it was written in the Ref. \cite{HaSh}.

\subsection{Levi-Civita-covariant shape of equations}

It is also possible to rewrite the equation of motion (\ref{fulleq}) in terms of the Levi-Civita covariant derivative of the tensor ${\mathfrak S}_{\mu\nu\alpha}$. Indeed, we have
$$\left(\bigtriangledown_{\alpha} + T_{\alpha} \right) {\mathfrak S}_{\mu}^{\hphantom{\mu}\nu\alpha} = {\mathop\bigtriangledown\limits^{(0)}}_{\alpha} {\mathfrak S}_{\mu}^{\hphantom{\mu}\nu\alpha}- K^{\beta}_{\hphantom{\beta}\alpha\mu}  {\mathfrak S}_{\beta}^{\hphantom{\beta}\nu\alpha} + K^{\nu}_{\hphantom{\nu}\alpha\beta}  {\mathfrak S}_{\mu}^{\hphantom{\mu}\beta\alpha} =  {\mathop\bigtriangledown\limits^{(0)}}_{\alpha} {\mathfrak S}_{\mu}^{\hphantom{\mu}\nu\alpha}+ K_{\alpha\beta\mu}  {\mathfrak S}^{\alpha\beta\nu} -\frac12 T^{\nu\alpha\beta}  {\mathfrak S}_{\mu\alpha\beta} , $$
for the difference between the two connections in serving the index ${}^{\alpha}$ has already been taken care of in terms of the torsion vector $T_{\alpha}$.

Altogether, we get a very nice form of equations for an arbitrary New GR model,
$$ {\mathop\bigtriangledown\limits^{(0)}}_{\alpha} {\mathfrak S}_{\mu}^{\hphantom{\mu}\nu\alpha}+ {\mathfrak S}^{\alpha\beta\nu}\left( K_{\alpha\beta\mu} -  T_{\alpha\beta\mu} \right)+ \frac12 {\mathfrak T}\delta^{\nu}_{\mu}=0,$$
the same as in TEGR \cite{meCov}, modulo the substitution of $\mathbb T$ and $S^{\mu\nu\alpha}$ by $\mathfrak T$ and ${\mathfrak S}^{\mu\nu\alpha}$. What was not noticed in the Ref. \cite{meCov} is that it can be brought to an even nicer form. Indeed, using the obvious (due to symmetry of the Levi-Civita connection) relation 
$$ T_{\alpha\beta\mu}= K_{\alpha\beta\mu} - K_{\alpha\mu\beta},$$
the equation (\ref{fulleq}) gets reduced to
\begin{equation}
\label{LCeq}
 {\mathop\bigtriangledown\limits^{(0)}}_{\alpha} {\mathfrak S}_{\mu}^{\hphantom{\mu}\nu\alpha}- {\mathfrak S}^{\alpha\nu\beta} K_{\alpha\mu\beta} + \frac12 {\mathfrak T}\delta^{\nu}_{\mu}=0.
\end{equation}
Note that until now we have been working purely in vacuum.

\subsection{Bianchi identities}

An important point to make is that the teleparallel models are invariant under diffeomorphisms. Indeed, the torsion tensor (\ref{tor}) is defined in terms of an antisymmetrised derivative (\ref{Weitz}) of the 1-forms $e^a_{\mu}$, and therefore transforms as a tensor under coordinate changes,  indeed. Hence, according to the second Noether theorem, the equations of motion satisfy an identical equality. The derivation goes with no difference in every torsion-based theory, and therefore for the proof we just refer the reader to the Section VI A of the Ref. \cite{HaSh} which proposed the New GR models or to the Ref. \cite{Bianchi} where it was presented in the context of $f(\mathbb T)$ gravity but with no use of the particular action structure in the procedure.

With the symbol ${\mathfrak T}^{\mu\nu}$ we denote the variation of the action with respect to the tetrad $e^a_{\nu}$, and with the lower tangent space index replaced by a coordinate one (by the usual action of the tetrad) and then raised, i.e. the quantity which stands in front of $g_{\mu\beta} e^{\beta}_a  \delta e^a_{\nu}$ in the variation. The generalised Bianchi identity \cite{Bianchi} is then
\begin{equation}
\label{Noether}
{\mathop{\bigtriangledown}\limits^{(0)}} {}_{\nu}  {\mathfrak T}^{\mu\nu} + K^{\alpha\mu\beta} {\mathfrak T}_{\alpha\beta}=0.
\end{equation}
Note that, due to antisymmetry of the contortion, $K_{\alpha\mu\beta}=-K_{\beta\mu\alpha}$, once the antisymmetric part of the equations is made zero, the symmetric part satisfies the usual Bianchi identity. Therefore, with no trouble we can couple ordinary matter, with symmetric covariantly-conserved energy momentum tensor ${\mathcal T}_{\mu\nu}$, to teleparallel theories of gravity:
\begin{equation}
\label{eqwmat}
 {\mathop\bigtriangledown\limits^{(0)}}_{\alpha} {\mathfrak S}_{\mu}^{\hphantom{\mu}\nu\alpha}- {\mathfrak S}^{\alpha\nu\beta} K_{\alpha\mu\beta} + \frac12 {\mathfrak T}\delta^{\nu}_{\mu}=\kappa {\mathcal T}_{\mu}^{\nu}.
\end{equation}
Moreover, any symmetric tensor which happened to serve as the right hand side (\ref{eqwmat}) is automatically required to be covariantly conserved due to the Bianchi identity (\ref{Noether}).

\section{Spherically symmetric configurations}

As we will comment more on below, for facilitating the search for exact solutions, it makes sense to write the metric without fixing a particular radial variable. It means using the expression
\begin{equation}
\label{themetr}
g_{\mu\nu}dx^{\mu}dx^{\nu}=A^2(r)dt^2-B^2(r)dr^2-h^2(r)(d\theta^2+sin^2(\theta) d\phi^2)
\end{equation}
for the metric which corresponds to the following Levi-Civita connection coefficients
$$\mathop\Gamma\limits^{(0)}{\vphantom{\Gamma}}^{1}_{00}=\frac{AA^{\prime}}{B^2}, \qquad \mathop\Gamma\limits^{(0)}{\vphantom{\Gamma}}^{1}_{11}=\frac{B^{\prime}}{B}, \qquad \mathop\Gamma\limits^{(0)}{\vphantom{\Gamma}}^{1}_{22}=-\frac{hh^{\prime}}{B^2},\qquad \mathop\Gamma\limits^{(0)}{\vphantom{\Gamma}}^{1}_{33}=-\frac{hh^{\prime}\sin^2\theta}{B^2},$$
$$\mathop\Gamma\limits^{(0)}{\vphantom{\Gamma}}^{2}_{33}=-\sin(\theta)\cos(\theta),  \qquad    \mathop\Gamma\limits^{(0)}{\vphantom{\Gamma}}^{0}_{10}=\frac{A^{\prime}}{A}, \qquad \mathop\Gamma\limits^{(0)}{\vphantom{\Gamma}}^{2}_{12}=\mathop\Gamma\limits^{(0)}{\vphantom{\Gamma}}^{3}_{13}=\frac{h^{\prime}}{h}, \qquad \mathop\Gamma\limits^{(0)}{\vphantom{\Gamma}}^{3}_{23}=\cot\theta.$$
The usual surface-area radius corresponds to the choice of $h(r)=r$.

These formulae are very standard, of course. However, the choice of a tetrad for representing the metric is also very important in teleparallel theories. There are many poor choices which would not allow us to solve the equations. And indeed, once the $f(\mathbb T)$ theories had surfaced, many works appeared with false results, taking a tetrad which could never be a solution and simply ignoring the antisymmetric part of equations.

The most common naive choice of the tetrad for the metric (\ref{themetr}) would be a diagonal one, diagonal in the very same spherical coordinates. It does not work, unless in pure TEGR. In modern literature, we have a slang of "good" and "bad" tetrads. We must say, it is not a very clean language, since this is just about satisfying or not satisfying a particular part of equations. Actually, it is nothing but very natural that a diagonal in spherical coordinates tetrad is no good, when the tetrad is a dynamical variable. The construction of spherical coordinates does not respect the symmetry and, moreover, it is singular at the polar line.

The "good" and symmetry-respecting tetrad\footnote{This tetrad is actually diagonal in coordinates of Cartesian type \cite{meSpher}, and therefore it is precisely the tetrad which has been used in the seminal paper \cite{HaSh}.}, which is quite standard by now, for the metric (\ref{themetr}) can be taken as
\begin{equation}
\label{tetr}
e^{a}_{\mu} = \left(
\begin{array}{cccc}
A(r) & 0 & 0 & 0\\
0 & B(r) \sin(\theta)\cos(\phi) & h(r) \cos(\theta) \cos(\phi) & -h(r)\sin(\theta) \sin(\phi) \\
0 & B(r) \sin(\theta) \sin(\phi) & h(r) \cos(\theta) \sin(\phi) & h(r) \sin(\theta) \cos(\phi) \\
0 & B(r) \cos(\theta) & -h(r)\sin(\theta) & 0
\end{array}
\right)
\end{equation}
which, as was shown in the Refs. \cite{Obukhov, meSpher}, allows one to easily find all the non-zero torsion tensor (\ref{tor}) components as
$$T^{0}_{\hphantom{0}10}=-T^{0}_{\hphantom{0}01}=\frac{A^{\prime}}{A}, \qquad T^{2}_{\hphantom{0}12}=T^{3}_{\hphantom{0}13}=-T^{2}_{\hphantom{0}21}=-T^{3}_{\hphantom{0}31}=-\frac{B- h^{\prime} }{h}$$
or upon lowering the index:
$$T_{010}=-T_{001}=AA^{\prime}, \qquad T_{212}=-T_{221}=h(B-h^{\prime}), \qquad T_{313}=-T_{331}=h(B-h^{\prime}) \cdot \sin^2 \theta,$$
and with the only non-zero torsion vector component being
$$T_1=\frac{A^{\prime}}{A} - 2\cdot \frac{B- h^{\prime} }{h}$$
where we have enumerated the coordinates $t,r,\theta,\phi$ as $0,1,2,3$ respectively.

Note that the only non-zero torsion tensor components are those with one index relating to the radius and the other two indices being equal to each other. We immediately see that the same is true of the contortion
$$K_{100}=-K_{001}=AA^{\prime}, \qquad K_{122}=-K_{221}=h(B - h^{\prime}), \qquad K_{133}=-K_{331}=h(B - h^{\prime}) \cdot \sin^2 \theta$$
and the superpotential (\ref{superpot})
$${\mathfrak S}_{010}=-{\mathfrak S}_{001}=\left(\frac{a+b}{2}-c\right)\cdot A A^{\prime} + 2c\cdot  \frac{A^2\left(B-h^{\prime}\right)}{h},$$ 
$${\mathfrak S}_{212}=-{\mathfrak S}_{221}=\left(\frac{a+b}{2}-2c\right)\cdot h \left(B-h^{\prime}\right) + c\cdot  \frac{h^2 A^{\prime}}{A},$$
$${\mathfrak S}_{313}=-{\mathfrak S}_{331}=\left(\left(\frac{a+b}{2}-2c\right)\cdot h \left(B-h^{\prime}\right) + c\cdot  \frac{h^2 A^{\prime}}{A}\right) \cdot \sin^2 \theta.$$ 
If to put  $a=b=c=1$, this is the standard superpotential tensor  of TEGR. Note also that only $c$ and $a+b$ have any relevance in all these expressions. It is a reflection of the fact that, due to all non-zero torsion tensor components having two indices equal, the axial-vector part of the torsion is identically zero for the chosen tetrad Ansatz (\ref{tetr}).

\section{Explicit equations and solutions in New GR}

Now we clearly have all the tensorial components we need, and for brevity of notation, we define
$$\overline{a}\equiv \frac{a+b}{2}.$$
It is very easy to calculate the torsion scalar (\ref{torscal})
\begin{equation}
\label{theT}
{\mathfrak T} = \left(c - \overline{a} \right)\cdot \frac{{A^{\prime}}^2}{A^2 B^2} +(4c-2\overline{a})\cdot \frac{(B-h^{\prime})^2}{h^2 B^2}-4c\cdot \frac{A^{\prime} (B- h^{\prime})}{hAB^2} 
\end{equation}
which, in the case of TEGR ($a=b=c=1$) and $h(r)=r$, coincides with the known result \cite{morespher, Bianchi, meSpher}. In order  to finally write down the equations (\ref{LCeq}) for the chosen tetrad (\ref{tetr}) in vacuum, the only slightly non-trivial, but pretty standard for any GR-practitioner task is to evaluate 
$$ {\mathop\bigtriangledown\limits^{(0)}}_{\alpha} {\mathfrak S}_{\mu}^{\hphantom{\mu}\nu\alpha}=\partial_{\alpha}{\mathfrak S}_{\mu}^{\hphantom{\mu}\nu\alpha} + \mathop\Gamma\limits^{(0)}{\vphantom{\Gamma}}^{\alpha}_{\alpha\rho}{\mathfrak S}_{\mu}^{\hphantom{\mu}\nu\rho} - \mathop\Gamma\limits^{(0)}{\vphantom{\Gamma}}^{\rho}_{\alpha\mu}{\mathfrak S}_{\rho}^{\hphantom{\mu}\nu\alpha}=\frac{1}{\sqrt{-g}}\partial_{\alpha}\left(\sqrt{-g}{\mathfrak S}_{\mu}^{\hphantom{\mu}\nu\alpha}\right) - \mathop\Gamma\limits^{(0)}{\vphantom{\Gamma}}^{\rho}_{\alpha\mu}{\mathfrak S}_{\rho}^{\hphantom{\mu}\nu\alpha}.$$ 
In particular, it is quite easy to check that there are no off-diagonal, let alone antisymmetric equations, for all their potential contributions are identically zero.

In the temporal component ${\mathfrak T}^0_0$, we find that $ \mathop\Gamma\limits^{(0)}{\vphantom{\Gamma}}^{\rho}_{\alpha 0}{\mathfrak S}_{\rho}^{\hphantom{\mu} 0\alpha} = {\mathfrak S}^{\alpha 0 \beta} K_{\beta 0 \alpha}$, and the temporal equation of motion takes the form of $\frac{1}{\sqrt{-g}}\partial_{\alpha}\left(\sqrt{-g}{\mathfrak S}_{0}^{\hphantom{\mu}0\alpha}\right)  + \frac12 {\mathfrak T}=0$:
\begin{equation}
\label{temp}
\left[\left(\overline{a}-c\right)\cdot \frac{A^{\prime}}{AB^2} + 2c\cdot \frac{B-h^{\prime}}{hB^2} \right]^{\prime} + \left(\frac{A^{\prime}}{A}+\frac{B^{\prime}}{B}+2\frac{h^{\prime}}{h}\right) \cdot \left[\left(\overline{a}-c\right)\cdot \frac{A^{\prime}}{AB^2} + 2c\cdot \frac{B-h^{\prime}}{hB^2}\right] + \frac12 {\mathfrak T}=0.
\end{equation}

In the radial component ${\mathfrak T}^1_1$, we immediately see that the only non-zero contribution, apart from $\mathfrak T$, is from the term of $-\mathop\Gamma\limits^{(0)}{\vphantom{\Gamma}}^{\rho}_{\alpha 1}{\mathfrak S}_{\rho}^{\hphantom{\mu}1 \alpha}$ which gives the equation
\begin{equation}
\label{rad}
\left(\overline{a}-c\right)\cdot \frac{{A^{\prime}}^2}{A^2 B^2} + 2c\cdot \frac{A^{\prime}(B-h^{\prime})}{hAB^2}+ (4c-2\overline{a})\cdot \frac{h^{\prime}(B-h^{\prime})}{h^2 B^2} -2c\cdot \frac{h^{\prime} A^{\prime}}{hAB^2} +\frac12 {\mathfrak T}=0.
\end{equation}

Finally, the angular component ${\mathfrak T}^2_2={\mathfrak T}^3_3$ does not have that much of cancellations, but the result (after flipping the overall sign) is also quite simple:
\begin{equation}
\label{ang}
\left[\left(\overline{a}-2c\right)\cdot \frac{B-h^{\prime}}{hB^2} + c\cdot \frac{A^{\prime}}{AB^2} \right]^{\prime} + \left(\frac{A^{\prime}}{A}+\frac{B^{\prime}}{B}+\frac{2h^{\prime}-B}{h}\right) \cdot \left[\left(\overline{a}-2c\right)\cdot \frac{B-h^{\prime}}{hB^2} + c\cdot \frac{A^{\prime}}{AB^2}\right] - \frac12 {\mathfrak T}=0.
\end{equation}

One can also check it here that, in the case of TEGR with $h(r)=r$, the equations (\ref{theT}, \ref{temp}, \ref{rad}, \ref{ang}) successfully reproduce what we already knew \cite{morespher, Bianchi, meSpher} in the $f(\mathbb T)$ models\footnote{ In the paper \cite{meSpher}, there is a misprint in the temporal equation (eq. (13) there): a forgotten square over the radial variable in the prefactor of the $f_T$ term.} when $f(\mathbb T)=\mathbb T$. Note that, due to the freedom of choosing the radial coordinate, there are only two independent functions to satisfy these three equations. They are non-trivially solvable only due to the Bianchi identity (\ref{Noether}). It can also be checked directly at the obtained equations, though the calculations would be quite cumbersome. On the other hand, in the case of $\overline{a}=c$ those are the same as GR equations in vacuum\footnote{In the case of $\overline{a}=c=1$ the equations are precisely those of GR, reflecting again the fact that the axial torsion is identically zero for our tetrad (\ref{tetr}), while any other non-zero value of $\overline{a}=c$ amounts to mere renormalisation of the gravitational constant which has no influence on vacuum solutions. }, therefore it is enough to check it for a single complementary case of, say, $c=0$ and $\overline{a}=1$.

\subsection{The choice of conformally-Euclidean spatial coordinates}

All the static spherically symmetric solutions can be found using the radial coordinate which naturally arises by coordinate change from the Cartesian coordinates of the Ref. \cite{HaSh}: $$h(r) = B(r) r.$$ 
The metric (\ref{themetr}) takes the form of
$$g_{\mu\nu}dx^{\mu}dx^{\nu}=A^2(r)dt^2-B^2(r) \left( dr^2+ r^2 d\theta^2 + r^2 sin^2(\theta) d\phi^2 \right),$$
and as we show below, all the equations can be put into a very nice form. These coordinates are often called isotropic, however their true meaning is that they represent the spatial slice in an explicitly conformally Euclidean form. Note also that, in these coordinates, we can freely change the overall normalisations of both $A$ and $B$ by rescaling the time and the radius with arbitrary non-zero constant factors.

We start from noticing that, having substituted this choice of $h(r)$, the radial equation (\ref{rad}) gets a term of $-{\mathfrak T}$ in its left hand side, and acquires an alternative form of
\begin{equation}
\label{confrad} 
\frac12 B^2 {\mathfrak T}=- 2c\cdot \frac{A^{\prime}}{rA} - (4c-2\overline{a})\cdot\frac{B^{\prime}}{rB}
\end{equation}
with the torsion scalar (\ref{theT})
\begin{equation}
\label{confT}
B^2{\mathfrak T}= (c-\overline{a})\cdot \frac{{A^{\prime}}^2}{A^2} +  (4c-2\overline{a})\cdot \frac{{B^{\prime}}^2}{B^2} + 4c\cdot  \frac{A^{\prime} B^{\prime}}{AB}.
\end{equation}
At the same time, the temporal (\ref{temp}) and the angular (\ref{ang}) equations turn into
\begin{equation}
\label{conftemp}
\left[\left(\overline{a}-c\right)\cdot \frac{A^{\prime}}{A} - 2c \cdot\frac{B^{\prime}}{B}\right]^{\prime} + \left(\frac{A^{\prime}}{A} + \frac{B^{\prime}}{B} + \frac{2}{r}\right) \cdot \left[\left(\overline{a}-c\right)\cdot \frac{A^{\prime}}{A} - 2c \cdot\frac{B^{\prime}}{B}\right] =-\frac12 B^2 \mathfrak T 
\end{equation}
and
\begin{equation}
\label{confang}
\left[c \cdot \frac{A^{\prime}}{A} + \left(2c - \overline{a}\right) \cdot\frac{B^{\prime}}{B}\right]^{\prime} + \left(\frac{A^{\prime}}{A} + \frac{B^{\prime}}{B} + \frac{1}{r}\right) \cdot \left[c \cdot \frac{A^{\prime}}{A} + \left(2c - \overline{a}\right) \cdot\frac{B^{\prime}}{B}\right]=\frac12 B^2 \mathfrak T 
\end{equation}
respectively.

We immediately see that substituting the new value of $\mathfrak T$ from the radial equation (\ref{confrad}) into the angular equation (\ref{confang}) one gets
$$\left(rAB\left[c \cdot \frac{A^{\prime}}{A} + \left(2c - \overline{a}\right) \cdot\frac{B^{\prime}}{B}\right]\right)^{\prime} = -2AB\left[c \cdot \frac{A^{\prime}}{A} + \left(2c - \overline{a}\right) \cdot\frac{B^{\prime}}{B}\right]. $$
Since the equation $f^{\prime}(r)=-\frac{2f(r)}{r}$ obviously requires $f(r)\propto \frac{1}{r^2}$, we find out that
\begin{equation}
\label{fw}
c \cdot \frac{A^{\prime}}{A} + \left(2c - \overline{a}\right) \cdot\frac{B^{\prime}}{B}=\frac{f_1}{ABr^3}
\end{equation}
with an arbitrary integration constant $f_1$.

With the temporal equation (\ref{conftemp}), it does not immediately go like that. One has to find a linear combination of the temporal  (\ref{conftemp}) and angular  (\ref{confang}) equations which again gets a derivative proportional to its argument. It's not difficult to deduce that adding twice the angular  equation (\ref{confang}) to the temporal  one (\ref{conftemp})  yields a relation which, by substituting the value of $\mathfrak T$ from the radial equation (\ref{confrad}) again, can be transformed into
$$\left(rAB\left[ \left(\overline{a} + c\right) \cdot \frac{A^{\prime}}{A} + \left(2c-2\overline{a}\right) \cdot\frac{B^{\prime}}{B}\right]\right)^{\prime} = -AB\left[ \left(\overline{a} + c\right) \cdot \frac{A^{\prime}}{A} + \left(2c-2\overline{a}\right) \cdot\frac{B^{\prime}}{B}\right]. $$
This time, the equation $f^{\prime}(r)=-\frac{f(r)}{r}$ demands $f(r)\propto \frac{1}{r}$, and we get
\begin{equation}
\label{sw}
\left(\overline{a} + c\right) \cdot \frac{A^{\prime}}{A} + \left(2c-2\overline{a}\right) \cdot\frac{B^{\prime}}{B}=\frac{f_2}{ABr^2}
\end{equation}
with another arbitrary constant $f_2$.

Assuming for now that ${\overline a} (3c - {\overline a})\neq 0$, we can take an obvious linear combination of the two derived equations (\ref{fw}) and (\ref{sw})
$$\frac{A^{\prime}}{A} + \frac{B^{\prime}}{B}=\frac{1}{\overline{a}(3c-\overline{a})AB}\left(\frac{f_1 (3\overline{a}-c)}{r^3}+\frac{f_2 (c-\overline{a})}{r^2}\right)$$
which gives an immediate solution for the function $AB$:
\begin{equation}
\label{AB}
AB=f_3 - \frac{f_1 (3\overline{a}-c)}{2\overline{a}(3c-\overline{a})}\cdot\frac{1}{r^2} - \frac{f_2 (c-\overline{a})}{\overline{a}(3c-\overline{a})}\cdot\frac{1}{r}.
\end{equation}
In case of asymptotically flat spacetimes, we need $f_3 \neq 0$ which can then be put to $f_3 =1$ by constant rescaling of time and/or radius.

Having done so, two other evident combinations of the equations (\ref{fw}) and (\ref{sw}) give us the final results for $A$ and $B$:
\begin{equation}
\label{A}
\frac{A^{\prime}}{A}= \frac{1}{\overline{a}(\overline{a}-3c)}\left(\frac{2(c-\overline{a})f_1}{ABr^3} - \frac{(2c-\overline{a})f_2}{ABr^2}\right),
\end{equation}
\begin{equation}
\label{B}
\frac{B^{\prime}}{B}= \frac{1}{\overline{a}(3c-\overline{a})}\left(\frac{(c+\overline{a})f_1}{ABr^3} - \frac{cf_2}{ABr^2}\right).
\end{equation}
Obviously, with the solution (\ref{AB}) for $AB$ and for any model parameters, these equations can be integrated in elementary functions. Moreover, it is enough to solve any one of them, and another function will be given by combination with $AB$.

The result is a bit too generic yet. Strictly speaking, in the formulae (\ref{A}) and (\ref{B}) supplied with the $AB$ function (\ref{AB}), we have found all the solutions of the temporal (\ref{conftemp}) and angular (\ref{confang}) equations, blindly assuming that the radial equation (\ref{confrad}) was satisfied, too. The assumption must also be checked. In principle, the three equations are not independent, due to the Bianchi identities (\ref{Noether}). However, the radial one (\ref{confrad}) is a constraint, in the sense of a lower derivative order, which must impose restrictions on the integration constants. 

Therefore, having obtained a solution of equations (\ref{A}) and (\ref{B}), one must substitute it into the radial equation (\ref{confrad}) and check which values of the integration constants are compatible with it. Actually, a reasonable way is to simply substitute the expressions  (\ref{A}) and (\ref{B}) for $\frac{A^{\prime}}{A}$ and $\frac{B^{\prime}}{B}$ with the formula (\ref{AB}), and then the radial equation (\ref{confrad}) yields an algebraic constraint on the integration constants $f_1$ and $f_2$. Note that the original New GR paper \cite{HaSh} went another way. They found the necessary relation between $f_1$ and $f_2$ by considering the weak field limit at $r\to\infty$ first. As we mentioned in the Introduction, the relation of the constants to the central mass is a matter of interpretation, however their relation to each other is an important restriction imposed by the equations.

Since we are not studying all these solutions in detail, we will not write their explicit general expressions here. Those are not very illuminating by themselves. Note though that, in the Ref. \cite{HaSh} it was assumed that the quadratic polynomial in the denominators of the formulae (\ref{A}) and (\ref{B}) does have two different real roots, which is always true when the model parameters are not too far from their values in the TEGR action, and explicit solutions in terms of those roots were presented, whose geometric properties were also  studied later \cite{ HaSh2}.

\subsection{An explicit example}

As a relatively simple example which goes beyond the solutions in terms of power-law functions only, in the classical reference \cite{HaSh}, let us take $c=3\overline{a}$. In this case, the solution (\ref{AB}) takes the form of
$$AB=f_3 - \frac{f_2}{4\overline{a} r}.$$
Then it is enough to solve the equations for any one of the two functions. Let us take the function $A$ and transform its equation (\ref{A}) to an easily integrable form:
$$\frac{A^{\prime}}{A}=\frac{5f_2 r - 4f_1}{2r^2 (4 \overline{a} r - f_2)} =\frac{2f_1}{f_2 r^2} + \frac{\left(\frac52 - \frac{8 \overline{a} f_1}{f_2^2}\right)\frac{f_2}{4\overline{a}}}{\left(r-\frac{f_2}{4\overline{a}}\right)r}$$
where we have assumed $f_3=1$. The case of $f_3=0$ is very easy to handle. But if we want an asymptotically flat solution, then $f_3\neq 0$ and, without loss of generality, we can assume that in our coordinates $f_3=1$ indeed.

However, before doing the job, let's substitute the equations (\ref{AB}, \ref{A}, \ref{B}) for our model into the radial equation (\ref{confrad}). The result
$$\frac{5f_2^2}{16 \overline{a} r^4}= \frac{2f_1 f_3}{r^4},$$
keeping in mind that $f_3=1$, allows us to simplify the equation for $A$ even a bit more:
$$\frac{A^{\prime}}{A}=\frac{5}{16 \overline{a} r^2} + \frac{\frac54 \cdot \frac{f_2}{4\overline{a}}}{\left(r-\frac{f_2}{4\overline{a}}\right)r}.$$
Using the obvious formula of $\left(\log \left(1-\frac{r_0}{r}\right)^n\right)^{\prime}=\frac{n r_0}{r(r- r_0)}$, we get the final result
$$A=\left(1-\frac{f_2}{4\overline{a} r}\right)^{\frac54} e^{-\frac{5}{16\overline{a}r}}, \qquad B=\frac{AB}{A}=\left(1-\frac{f_2}{4\overline{a} r}\right)^{-\frac14} e^{\frac{5}{16\overline{a}r}},$$
see also the Ref. \cite{Obukhov}.

Obviously, the would-be horizon is much less innocent than its counterpart in GR. It has been noticed already in the classical paper \cite{HaSh2} that, even from the point of view of purely metric geometry, generically the New GR solutions have singular horizons. We are not going into a deeper discussion of the corresponding geometry \cite{Obukhov} in this paper. However, we would also abstain from saying that there are no Black Holes in these theories \cite{Obukhov}. In a sense, the singular horizons might be taken similarly to the popular firewalls.

Note, at the same time, that even the usual Schwarzschild horizon is anyway singular in terms of the tetrad Ansatz (\ref{tetr}), or its corresponding torsion \cite{Obukhov}, as was also mentioned, for example, in the Ref. \cite{notspher}. And it is very natural indeed. One of the features of the teleparallel geometry (\ref{tetr}) is that the normalised parallelly-transported time-like vector $e_0$ turns into a space-like one when crossing the horizon. Of course, it cannot be in any way smooth.

\subsection{The special cases}

Coming back to the special cases which were excluded above, let's first assume 
$$\overline{a}=0.$$ 
Then both equations (\ref{fw}) and (\ref{sw}) get $\frac{A^{\prime}}{A} +2 \frac{B^{\prime}}{B}$ in their left hand sides. Since it is impossible to simultaneously have two different powers of $r$ in expressions for one and the same quantity, we must take 
$$\frac{A^{\prime}}{A} +2 \frac{B^{\prime}}{B}=0$$
and check it with the radial equation (\ref{confrad}). An amazing result is that ${\mathfrak T}=0$ and the radial equation (\ref{confrad}) is identically satisfied.

Another special option is
$$\overline{a}=3c$$
with the equations (\ref{fw}) and (\ref{sw}) showing $\frac{A^{\prime}}{A} - \frac{B^{\prime}}{B}$ being equal to two different functions again, and therefore requiring
$$\frac{A^{\prime}}{A} - \frac{B^{\prime}}{B}=0$$
with the same result of the  radial equation (\ref{confrad}) check: ${\mathfrak T}=0$ and all the equations are satisfied.

All in all, we have found two cases of degeneracy in the solutions: the model of $\overline{a}=0$ is solved by an arbitrary function $B$ as long as $A\propto B^{-2}$, and the same is true of the model of $\overline{a}=3c$ with $A\propto B$. The former is the case of any model which depends on torsion vector and torsion axial vector only, while the latter is a conformally flat spacetime with a conformal factor being an arbitrary function of radius. 

To summarise this finding, we again see that different New GR models behave in genuinely different ways. At the level of static spherically symmetric solutions, in these special cases an enhanced freedom of solutions arises, while a model with a Lagrangian depending on the axial torsion only ($c=0$ and $b=-a$) would have an absolutely arbitrary tetrad of this form (\ref{tetr}) as a solution. Analogously, differences can be found in the Hamiltonian analysis of various New GR models \cite{Dan1, Shy, Dan2}.

\section{The GR solutions}

To reproduce the vacuum solutions of GR, we have to take $c=\overline{a}$. And let's assume for simplicity that $c=\overline{a}=1$. Otherwise, the renormalised gravitational constant can be absorbed into the integration constants below. In this case, the angular equation (\ref{fw})
$$\frac{A^{\prime}}{A} + \frac{B^{\prime}}{B}=\frac{f_1}{ABr^3}$$
immediately gives the function $AB$:
$$AB= f_3 - \frac{f_1}{2r^2} = 1 -\frac{M^2}{4r^2}$$
where we chose $f_3=1$ and denoted $f_1\equiv \frac{M^2}{2}$, while our mixed equation (\ref{sw})
$$2\frac{A^{\prime}}{A}= \frac{f_2}{ABr^2}$$
gives then the final result.

If we forget to check it with the radial equation (\ref{confrad}), then we solve the $2\frac{A^{\prime}}{A}=\frac{f_2 }{r^2-\frac{M^2}{4}}$ equation and get too many solutions of $A=f_4 \left(\frac{1-\frac{M}{2r}}{1+\frac{M}{2r}}\right)^{\frac{f_2}{2M}}$. However, if we do substitute $\frac{A^{\prime}}{A}$, $\frac{B^{\prime}}{B}$, and $AB$ into the radial equation (\ref{confrad}), we find out the necessary restriction on the integration constants
$$\frac{f_2^2- 8 f_1 f_3}{2r^4}=0.$$
In case of asymptotically flat solutions, we can rescale the coordinates in such a way as both $A$ and $B$ tend to unity at infinity, and therefore $f_3=1$. Then we get $f_2=\sqrt{8f_1}= 2M$, and the solution is
\begin{equation}
\label{Swarz}
A(r)=\frac{1-\frac{M}{2r}}{1+\frac{M}{2r}}, \qquad  B(r)=\frac{AB}{A}=\left(1+\frac{M}{2r} \right)^2.
\end{equation}

What we've got (\ref{Swarz}) is nothing but the Schwarzschild solution in isotropic coordinates. For the reader's convenience, let us briefly show how to see that. We want to rewrite 
$$A^2(r) dt^2 - B^2 (r)\left(dr^2 + r^2 \left(d\theta^2 + \sin^2(\theta) d\phi^2 \right) \right)\qquad  \mathrm{as} \qquad A^2(R) dt^2- {\tilde B}^2(R) dR^2 - R^2 \left(d\theta^2 + \sin^2(\theta) d\phi^2 \right).$$
From $R=\left(1+\frac{M}{2r} \right)^2 r$ we get $dR=\left(1-\frac{M^2}{4r^2}\right) dr$, and then ${\tilde B} dR= B dr=\frac{1+\frac{M}{2r}}{1-\frac{M}{2r}}dR$ implies ${\tilde B}=\frac{1}{A}$. On the other hand, $A^2=\frac{\left(1-\frac{M}{2r}\right)^2}{\left(1+\frac{M}{2r}\right)^2} =\frac{r}{R}\cdot \left(\left(1+\frac{M}{2r}\right)^2-\frac{2M}{r}\right)= 1- \frac{2M}{R}$ which finishes the proof.

Note that generically it won't be so easy to transform one formula into another. In the presentation above we avoided  finding $r=r(R)$ explicitly, even though it wouldn't be too difficult. In more complicated situations it might be even impossible in terms of  functions which are elementary enough, let alone being nice and compact in writing. A lesson to take is that in search for such solutions one should attempt using different possible radial variables. In particular, we probably need to try other options in our long, but largely unsuccessful, go for constructing these configurations in $f(\mathbb T)$ models \cite{notspher, morespher}.

On the other hand, it is important to note that the solutions are different from the general relativistic ones only if ${\overline a}\neq c$, or in our initial notation $a+b \neq 2c$. Even with irrelevance of axial torsion, it still leaves two parameters free, one of which is just normalisation of the effective gravitational constant though. In the Ref. \cite{HaSh} the latter parameter was disregarded and only one parameter family was left, from the viewpoint of preserving the Newtonian limit. Nevertheless, any other case can generically be presented as one of theirs with a rescaled gravitational constant. 

At the same time, in the recent papers \cite{str1, str2}, people study the very same New GR theory using a very different approach. From the erroneous claim\footnote{They do not give enough details of how they derived this condition from the conservation laws. However, note that an ideal fluid approximation requires more than the symmetry itself, namely the equality of the radial and tangential pressures. In order to check the radial component of the Bianchi identities (\ref{Noether}) in the surface-area coordinates in terms of the energy and pressure, one has to take the radial pressure $p_r$ from the radial equation and the tangential pressure $p_t$ from the angular equation, and then to check whether $p^{\prime}_r +\frac{A^{\prime}}{A} (\rho+p_r)+\frac{2}{R} (p_r - p_t)=0$.} that energy-momentum conservation requires $a+b=2c$, they restricted the model parameters to only the ones which give precisely the same static spherically symmetric solutions as in GR. In other words, they only allowed for a change in the gravitational constant which was excluded in the Ref. \cite{HaSh}, but removed all other modifications of gravity which were the subject of Refs. \cite{HaSh, HaSh2}. Of course, even in these models new effects will appear when studying perturbations around, so that the axial torsion becomes active again. However, it is important to understand that at the background level there cannot be anything new compared to GR, except of a total rescaling of the gravitational constant, once and for all.

\section{An unusual solution\\
\small{An infinite staircase in a small room}}

\epigraph{Here they started climbing some wide steps, and Margarita began to think there would be no end to them. She was struck that the front hall of an ordinary Moscow apartment could contain this extraordinary invisible, yet quite palpable, endless stairway.}{Mikhail Bulgakov. Master and Margarita.\\ (English translation by Richard Pevear and Larissa Volokhonsky, Penguin Books 1997)}

As an example of an unusual solution, let us look at what can be found with $A=1$ in a model of New GR. The equations (\ref{fw}) and (\ref{sw}) then prescribe $\frac{B^{\prime}}{B}$ be equal to two different functions. The only way to get something different from Minkowski metric is to assume  either  ${\overline a} = c$ or  ${\overline a} = 2c$. The former is the case of GR, and we know that nothing interesting can be found there, therefore we take a model with
$${\overline a} = 2c.$$
Assuming the asymptotic flatness, the equation (\ref{sw}) then requires $B(r) = 1 + \frac{f}{r}$ with an integration constant $f$. Checking the radial equation (\ref{confrad}), we see that
$$A(r)=1, \qquad B(r) = 1 + \frac{f}{r}$$
is indeed a solution, and actually the one with $\mathfrak T =0$ again.

Treating it analogously to what was done for the Schwarschild spacetime above, one can find that for the surface-area radius $R=r+f$ we get ${\tilde B}(R) = \frac{1}{1-\frac{f}{R}}$ and the metric can be written as
\begin{equation}
\label{MM}
g_{\mu\nu}dx^{\mu}dx^{\nu}=dt^2- \frac{dR^2}{\left(1-\frac{f}{R}\right)^2}-R^2(d\theta^2+\sin^2(\theta) d\phi^2).
\end{equation}
With an arbitrary integration constant $f$, it is a solution of the model with ${\overline a} = 2c$, i.e. $a+b = 4c$.

For a usual observer, there is no gravity, however the spatial geometry has a very interesting shape. At the $R\to\infty$ infinity the space looks pretty much like a simple Minkowski. However, $R\to f$ is yet another infinity which cannot be reached in a finite time. Globally, the spatial topology is of ${\mathbb R}\times {\mathrm S}^2$ cylinder type. When coming to the centre of this frozen Universe, the areas of spheres stop decreasing properly, and the lucky inhabitant comes into an infinite volume inside a finite fence. 

\section{Conclusion}

We have presented a detailed introduction to the New GR models and their static spherically symmetric solutions in vacuum. Due to the diffeomorphism invariance and the corresponding Bianchi identities, these solutions exist for any choice of the radial variable. Moreover, in the so-called isotropic coordinates they all can be found in elementary functions. 

The models with $a+b=0$ and $a+b=6c$ have degeneracy in their equations and allow for infinite families of solutions parametrised by an arbitrary function of the radius. The model with $a+b=4c$ enables us to construct an empty cold world with an infinite-volume room for devil's parties. According to these examples, surprisingly and interestingly enough, the cases of ${\mathfrak T}=0$ appear to be special, akin to what is know and quite obvious in $f(\mathbb T)$ models.

An important point we would like to stress once more is that the isotropic, or conformally-Euclidean, coordinates appeared to be very convenient for finding the solutions. Needless to say, trying to use various radial variables might be of interest for analogous tasks in other modified gravity models, too.

{\bf Acknowledgments.} The Authors are grateful to Yuri Obukhov. When the first version of this work had appeared, he kindly brought his very interesting paper \cite{Obukhov} to our attention.

\end{document}